\documentclass[12pt,preprint]{aastex}

\begin{document}

\title{Orbital Migration of Protoplanets in a Marginally Gravitationally
Unstable Disk}

\author{Alan P.~Boss}
\affil{Department of Terrestrial Magnetism, Carnegie Institution, 
5241 Broad Branch Road, NW, Washington, DC 20015-1305}
\email{boss@dtm.ciw.edu}

\begin{abstract}
 
 Core accretion and disk instability require giant protoplanets to form 
in the presence of disk gas. Protoplanet migration models generally assume 
disk masses low enough that the disk's self-gravity can be neglected. 
However, disk instability requires a disk massive enough to be marginally
gravitationally unstable (MGU). Even for core accretion, a FU Orionis
outburst may require a brief MGU disk phase. We present a new set of 
three dimensional, gravitational radiation hydrodynamics models of MGU 
disks with multiple protoplanets, which interact gravitationally 
with the disk and with each other, including disk gas mass accretion. 
Initial protoplanet masses are 0.01 to 10 $M_\oplus$ for core accretion 
models, and 0.1 to 3 $M_{Jup}$ for Nice scenario models, starting on 
circular orbits with radii of 6, 8, 10, or 12 AU, inside a 0.091 $M_\odot$ 
disk extending from 4 to 20 AU around a $1 M_\odot$ protostar. Evolutions 
are followed for up to $\sim$ 4000 yr and involve phases of relative 
stability ($e \sim$ 0.1) interspersed with chaotic phases ($e \sim$ 0.4) 
of orbital interchanges. The 0.01 to 10 $M_\oplus$ cores can orbit stably 
for $\sim$ 1000 yr: monotonic inward or outward orbital migration of 
the type seen in low mass disks does not occur. A system with giant planet
masses similar to our Solar System (1.0, 0.33, 0.1, 0.1 $M_{Jup}$) was 
stable for over 1000 yr, and a Jupiter-Saturn-like system was stable for 
over 3800 yr, implying that our giant planets might well survive a MGU 
disk phase. 

\end{abstract}


\section{Introduction}

 The discovery of short-period giant planets forced theorists to study
inward orbital migration of giant planets formed at much greater distances. 
Attention has focused on Type I and Type II migration (Kley \& Nelson 2012),
the former dealing with $\sim 10 M_\oplus$ cores that migrate rapidly due 
to tidal torques with the gaseous disk, and the latter dealing with
$\sim M_{Jup}$ protoplanets that are massive enough to open a gap in
the disk, and thereafter evolve along with the disk. Unchecked inward
Type I migration presumably can lead to a loss of the protoplanet. 
However, in the core accretion scenario for giant planet formation,
inward Type I migration can speed the growth of a core by the sweeping 
up of the smaller bodies it encounters (e.g., Alibert et al. 2005),
albeit at the price of a free parameter that reduces the Type I migration
rate to a favorable value. Most Type I migration models consider disks
with masses low enough that the disk's self-gravity can be ignored.
However, analytical (Pierens \& Hur\'e 2005) and two-dimenensional  
hydrodynamical (Baruteau \& Masset 2008) studies of Type I migration in 
self-gravitating disks found that the inclusion of disk self-gravity
could lead to a significant reduction in the Type I migration rate.
Similarly, Nelson \& Benz (2003) found that even massive planets
undergoing Type I migration in gravitationally stable disks had their
migration rates reduced when the disk's self-gravity was included.
Disk instability scenarios for giant planet formation in self-gravitating
disks may sidestep the danger of Type I migration, as the clumps initially 
formed have masses of order 1 $M_{Jup}$ (e.g., Boss 2005), large enough to
open disk gaps and undergo Type II migration in a low mass disk.

 Marginally gravitationally unstable (MGU) disks are of interest from
several points of view. Solar-type young stars are observed to
undergo FU Orionis outbursts (e.g., Hartmann \& Kenyon 1996), where mass
accretion rates onto the central protostar increase dramatically and remain 
high for periods of order 100 yr. Such outbursts may well occur every $\sim 10^4$
yr in T Tauri stars. A leading explanation for FU Orionis outbursts
is a MGU disk (e.g., Zhu et al. 2010; Vorobyov \& Basu 2010). MGU disks
also offer an attractive mechanism for achieving the large-scale transport,
both inward and outward, of small particles in the solar nebula that
appears to be required to explain the presence of refractory grains
in comets (e.g., Brownlee et al. 2006; Boss et al. 2012), and for
mixing initially spatially heterogeneous distributions of isotopes
(Boss 2012a). Finally, MGU disks are required for the disk instability 
mechanism of giant planet formation to operate (e.g., Boss 1997).
Recent extremely high spatial resolution models have shown that disk
instability is able to produce fragments (even inside $\sim$ 10 AU) in MGU disks
with much longer cooling times than had previously been thought to be needed
(Pardekooper 2012; Meru \& Bate 2012), in strong support of the disk 
instability mechanism.

 Core accretion and disk instability both require giant protoplanets 
to form in the presence of the disk gas. Considerable efforts have
gone into theoretical studies of protoplanetary orbital migration
(reviewed by Kley \& Nelson 2012), yet nearly all models of the 
interactions of protoplanets with disk gas assume a disk mass low enough 
that the disk's self-gravity can be neglected, as previously noted.
The exceptions also include works by Boss (2005), Baruteau et al. (2011), 
and Michael et al. (2011), who all studied quite different initial 
conditions for MGU disks, and as a result found a wide range 
of outcomes, from large-scale inward orbital migration,
to relatively little orbital migration. 
 
 The Nice model has become a leading explanation for the orbital
evolution of the giant planets in our solar system (Tsiganis et al. 
2005; Gomes et al. 2005; Levison et al. 2008). In the Nice model, 
Saturn forms with an orbital period less than twice that of Jupiter,
but as both planets interact with a massive residual disk of 
planetesimals (following dissipation of the gaseous disk), their orbits
cross a 2:1 mean motion resonance, where Saturn's orbital period
equals twice that of Jupiter. At that point, their orbits are
destabilized, and Saturn undergoes a phase of close encounters 
with Uranus and Neptune, which are assumed to have formed outside
Saturn's orbit. The two ice giants are kicked further outward to their 
present orbits, while the two gas giants are left behind on slightly 
eccentric orbits, with $e \sim 0.05$ to 0.1. While the Nice model
was derived in the context of the core accretion model for giant
planet formation, the question arises as to the orbital stability
of multiple giant planet systems during a MGU phase, such as
during a FU Orionis outburst prior to gaseous disk dissipation in
the core accretion scenario, or as a result of formation by the 
disk instability mechanism.

 Boss (2005) studied the orbital evolution of single ``virtual 
protoplanets'' (VP) with initial masses of 1 $M_{Jup}$ embedded in 
a MGU disk. Here we present two new set of models, each with up to 
four VPs initially present in the MGU disk. In the first set, the VP 
masses are chosen to investigate the orbital evolution of $\sim$ Earth-mass
cores trying to accrete gas during an MGU disk phase, and in the second set,
to investigate the evolution of already formed giant planets embedded 
in MGU disks, a situation analogous to the Nice model of giant 
planet evolution in gas-free, massive planetesimal disks.

\section{Numerical Methods for the Disk}

 The three dimensional (3D) numerical hydrodynamics code used is the same 
as that used in previous studies of disk instability (e.g., Boss 2005,
2010, 2011, 2012b), any one of which may be consulted for a brief summary
of the numerical techniques. A full description of the code and various
test cases is given by Boss \& Myhill (1992). The code has been also 
been tested specifically for accuracy in disk instability calculations 
(e.g., Boss 2012b, and references therein). 
 
 Compared to the models presented by Boss (2005), the only differences are 
that the equations were solved on a spherical coordinate grid with 
$N_\phi = 256$ and the number of terms in the spherical harmonic expansion 
for the gravitational potential of the disk was $N_{Ylm} = 32$. 
As in Boss (2005), the equations were solved with $N_r = 101$ and 
$N_\theta = 23$ in $\pi/2 \ge \theta \ge 0$. The radial grid was uniformly 
spaced with $\Delta r = 0.16$ AU between 4 AU and 20 AU. The $\theta$ 
grid was compressed into the midplane to ensure adequate vertical 
resolution ($\Delta \theta = 0.3^o$ at the midplane). While this spatial
resolution is sufficient to model the large-scale evolution of a MGU
disk, it may not be fine enough to properly resolve the Roche lobes and 
Hill spheres of individual protoplanets. However, given that the masses
of the even the most massive protoplanets that form are much less
than the total disk mass, to first order MGU disks evolve on their own,
with only minor perturbations from the embedded protoplanets. In fact,
searches for features in the disk gas distribution associated with
the more massive protoplanets did not reveal any clear structures.

 The Jeans length criterion (e.g., Truelove et al. 1997;
Boss 2002) was used to ensure that any clumps that formed were not 
numerical artifacts during the 400 yr of disk evolution leading from
the analytical initial conditions (see below) to the relaxed disk
phase when the protoplanets were inserted into the models. In the 
interests of pushing the protoplanet models as far as possible in time,
however, the Jeans length criterion was thereafter ignored, as it 
might at times have forced a significant refinement in the grid structure, 
slowing the subsequent evolution. This approach seemed reasonable,
as studying any subsequent clump formation was not the goal of 
these models. Even still, the models typically required from two to 
three years of continual computation on a dedicated cluster node.

\section{Initial Conditions for the Disk}

 The MGU disk system initially consists of a $1 M_\odot$ central protostar 
surrounded by a protoplanetary disk with a mass of 0.091 $M_\odot$ between 
4 AU and 20 AU. The initial protoplanetary disk structure 
is the same as that defined in Boss (2005), which is an
an approximate vertical density distribution (Boss 1993) for
an adiabatic, self-gravitating disk of arbitrary thickness in
near-Keplerian rotation about a point mass $M_s$

$$ \rho(R,Z)^{\gamma-1} = \rho_o(R)^{\gamma-1} $$
$$ - \biggl( { \gamma - 1 \over \gamma } \biggr) \biggl[
\biggl( { 2 \pi G \sigma(R) \over K } \biggr) Z +
{ G M_s \over K } \biggl( { 1 \over R } - { 1 \over (R^2 + Z^2)^{1/2} }
\biggr ) \biggr], $$

\noindent where $R$ and $Z$ are cylindrical coordinates,
$\rho_o(R)$ is the midplane density, and $\sigma(R)$ is the
surface density. The adiabatic pressure used in the initial model  
is defined by $p = K \rho^\gamma$, where the adiabatic constant 
$K = 1.7 \times 10^{17}$ (cgs units) and $\gamma = 5/3$ for the 
initial model. The radial
variation of the midplane density is a power law that ensures
near-Keplerian rotation throughout the disk

$$\rho_o(R) = \rho_{o4} \biggl( {R_4 \over R} \biggr)^{3/2}, $$

\noindent where $\rho_{o4} = 1.0 \times 10^{-10}$ g cm$^{-3}$ and
$R_4 = 4$ AU. A lower density halo $\rho_h$ of infalling molecular 
cloud gas and dust surrounds the disk, with

$$ \rho_h(r) = \rho_{h4} \biggl( {R_4 \over r} \biggr)^{3/2}, $$

\noindent where $\rho_{h4} = 1.0 \times 10^{-14}$ g 
cm$^{-3}$, and $r$ is the spherical coordinate radius. 

The initial temperature profile is based on the two dimensional 
radiative hydrodynamics calculations of Boss (1996), specifically
the ``standard model'' shown in Figure 9 of Boss (1996). The models have an
outer disk temperature of $T_o = 50$ K, resulting in an initial
$Q$ gravitational stability parameter as low as $Q_{min} = 1.5$ in the 
outermost disk, so that the outermost disk is gravitationally unstable. 
The initial midplane disk temperature at 4 AU (the inner boundary) 
is 600 K, leading to $Q > 10$ in the innermost disk and gravitational 
stability. Overall, then, the disk is marginally 
gravitationally unstable. The initial disk model is then
evolved for 400 yr ($3.8 \times 10^5$ time steps) before the 
protoplanets are inserted into the disk, in order to allow the 
disk to settle into a steady phase of disk instability, as shown
in Figure 1. Several distinct clumps and spiral arms are
evident at this initial time for the protoplanet evolutions,
allowing the models to investigate a ``worst case scenario'' for
the survival of protoplanets during an MGU disk phase.

\section{Numerical Methods for the Protoplanets}

 The protoplanets are handled in the same manner as described by
Boss (2005), where a dense clump was represented by a virtual 
protoplanet (VP) in the dynamically evolving disk. A VP is a point 
mass object that accretes mass and angular momentum from the disk, 
thereby determining its orbital evolution, subject to the gravitational 
forces of the central protostar and the spiral arms and clumps of 
the MGU disk, while the disk itself reacts to the gravitational 
force of the virtual protoplanet (VP). Rice et al. 
(2003) used a similar technique in their smoothed particle 
hydrodynamics (SPH) models of fragmentation in protostellar disks.
Krumholz et al. (2004) described their own technique for inserting
sink particles into an adaptive mesh refinement hydrodynamics code.

 Each VP is assumed to accrete mass $\dot M$ at the rate given by the 
Bondi-Hoyle-Lyttleton (BHL) formula (Livio 1986; Ruffert \& Arnett 1994)

$$ \dot M = { f 4 \pi \rho (G M)^2 \over (v^2 + c_s^2)^{3/2}}, $$

\noindent
where $f$ is a dimensionless coefficient, $G$ is the gravitational
constant, $M$ is the VP mass, $\rho$ is the local disk gas density, $c_s$
is the local sound speed, and $v$ is the speed of the VP through the
local gas. The VPs also accrete orbital angular momentum from the disk gas,
by accreting an amount of momentum from the local hydrodynamical
cell proportional to the mass being accreted from that cell, i.e., by
``consistent advection'', in such a way as to guarantee the conservation
of the total orbital angular momentum of the entire system. The mass 
and angular momentum accreted by each VP are removed from the cells in 
which they reside during the time step under consideration.

 The protoplanets' resulting updated velocity is used to calculate their
updated positions, to second-order accuracy in space and time, consistent 
with the accuracy of the hydrodynamical solution. To keep the system
synchronized, the same time step is used for both the disk hydrodynamics 
and the protoplanet orbital evolutions. The hydrodynamical time steps 
used are quite small, typically $< 10^{-3}$ yr. This is about 
$10^{-4}$ of an orbital period at 4 AU, and about $10^{-5}$ of an
orbital period at 20 AU. These exceedingly small time steps help to 
ensure the accuracy of the protoplanets' orbital evolutions; 
symplectic integrators typically achieve excellent energy conservation
when using time steps of order $10^{-3}$ the orbital period
(e.g., Chambers 2003).

 As in Boss (2005), the VPs affect the disk's evolution through having 
the gravitational potentials of each of the point masses ($- G M / R$) 
added into the total gravitational potential of the entire system, 
where the radius $R$ is the distance from a VP position 
to a cell center. This radius $R$ is constrained to be no smaller
than $\Delta r/2$, where $\Delta r$ is the local grid spacing
in the radial coordinate of the hydrodynamical grid, thereby softening
the gravitational potential in order to avoid singularities when
a VP approaches the center of a grid cell. The VPs evolve
as a result of the mass and angular momentum accreted, subject
to the gravitational potential of the protostar and disk, as well as to
the effects of centrifugal force. Gas drag is neglected, as is appropriate 
for these relatively massive protoplanets (cf. Boss et al. 2012, where 
gas drag is included for small particles evolving in MGU disks).
Numerical tests with $\dot M = 0$ and the gravitational potential of 
only a central protostar (Boss 2005) showed that VPs on initially
circular or elliptical orbits with semimajor axes of 5 AU
orbit stably for at least 500 years, with the VP's angular momentum 
being conserved to at least 8 digits. 

\section{Initial Conditions for the Protoplanets}

 For the $\sim$ Earth-mass protoplanet models, all four models
were initialized in the same manner, with the only variation
being in the initial mass of the protoplanet, as follows:
model a: 0.01 $M_\oplus$, model b: 0.10 $M_\oplus$,
model c: 1.0 $M_\oplus$, and model d: 10.0 $M_\oplus$. In
each case, four initially equal mass protoplanets were inserted in
the midplane locations denoted in Figure 1, at orbital
radii of 6 AU, 8 AU, 10 AU, and 12 AU, respectively, with the
variations in azimuthal locations evident in Figure 1. Each
protoplanet was inserted at the center of a hydrodynamical
cell with the same radial and azimuthal velocity as that of
the disk gas in that cell. Because the hydrodynamical grid
is restricted to the top hemisphere ($\pi/2 \ge \theta \ge 0$)
of the spherical coordinate grid, the protoplanets must be
limited to orbiting in the disk midplane. The $\sim$ Earth-mass
models were allowed to accrete mass according to the BHL formula
with $f = 10^{-4}$ being a factor in the formula. 
Nelson \& Benz (2003) explored a range of values of $f$, 
from 1 to $10^{-4}$. The coefficient $f$ should be less than unity because
of various effects neglected in the analysis, such as the accretion of
rotating gas of nonuniform density and temperature, shock fronts, 
and other effects as well, such as three dimensionality (e.g., 
Krumholz et al. 2005; Blondin \& Raymer 2012). Krumholz et al. 
(2005) found that $f$ could be as low as $10^{-2}$ purely
as a result of the vorticity of the gas being accreted.

 Table 1 summarizes the initial conditions and outcomes for the $\sim$ 
Earth-mass protoplanets, while Tables 2 and 3 show the same information 
for the giant protoplanet models, starting with VP masses in the range of
0.1 $M_{Jup}$ to 3 $M_{Jup}$, again at initial distances of 
6 AU, 8 AU, 10 AU, or 12 AU from the protostar. These models
were run with either $f = 10^{-4}$ or $10^{-3}$ in the BHL mass
accretion formula. Also shown in all three tables are the final masses
of the protoplanets ($M_f$) at the final time ($t_f$) for the
protoplanets that were still between 4 AU and 20 AU at the
end of the calculation. Protoplanets that hit either the inner
or the outer boundary are noted as having been ejected ``in'' or 
``out'', respectively. Note that close encounters between the
protoplanets need not lead to ``ejections'', so long as the planet
manages to stay between 4 AU and 20 AU. Similarly, not all
ejections need be the immediate result of a close encounter --
further interactions with the massive disk can also result in
a protoplanet hitting either the inner or outer disk boundary. In
any case, these ejections do not mean that the ejected protoplanet
has received enough of a kinetic energy increase to reach the
escape speed from the protostar and protoplanetary disk system.
In fact, because of the softening of the gravitational potential
of the protoplanets, close encounters between protoplanets cannot
have effectively closer approaches than $\Delta r/2 = 0.08$ AU,
and so cannot lead to protoplanets that reach the escape speed,
which would require closer approaches at least ten times smaller.
The models also do not include the effects of extended
atmospheres around the protoplanets, which could play a role
in very close encounters. Finally, the tables also list
the amount of disk mass that was accreted by the central protostar
during the evolutions, $\Delta M_s$.

\section{Results}

\subsection{Earth-Mass Protoplanets}

 We first consider the possible fates of $\sim$ Earth-mass 
cores that are attempting to accrete gas and become giant planets 
during a MGU disk phase. This question has not been considered to
date in the context of the classic core accretion scenario, where the
disk mass is assumed to be low enough to preclude gravitational
instability (e.g., Hubickyj et al. 2005; Lissauer et al. 2009)

 Figures 2 and 3 show the evolutions of the four $\sim$ Earth-mass 
protoplanet models. Monotonically inward migration of the type associated 
with Type I migration is not seen. In general, the protoplanets experienced
a significant amount of orbital perturbations driven by the clumps and
spiral arms in the MGU disk, resulting in continually evolving,
quasi-periodic changes in the orbital semimajor axes and eccentricities.
The time scales for these quasi-periodic wobbles in $a$ and $e$ are 
of order $\sim$ 30 yr, i.e., the orbital period at a distance of order
$\sim$ 10 AU, where the MGU disk is most active at forming transient
clumps and spiral arms (Figure 1). While vigorous, these perturbations
tend to average out to result in little net overall migration of the 
cores. Nevertheless, in each of the four models, at least one core
was perturbed enough to hit either the 4 AU or the 20 AU disk boundary; 
nearly half (7/16) of the cores were considered to be ``ejected'' in these
models (Table 1). Note that this relatively large fraction of ``ejected''
cores is in reality a severe overestimate, as in a more realistic
disk model, cores would not be considered lost unless they collided with
the central protostar, or were physically ejected from the system on
hyperbolic orbits. Given this caveat, the models depicted in Figure 2
show that $\sim$ Earth-mass cores should be able to survive a brief
($\sim 10^3$ yr) MGU phase of the sort associated with FU Orionis
outbursts, though a few cores might undergo large excursions in
semimajor axis as a result, and the surviving cores are likely to
be left on significantly eccentric ($e \sim 0.3$) orbits (Figure 3).
These results are relatively independent of the initial core
mass: all four models shown in Figures 2 and 3 appear qualitatively
similar.

 Because the Earth-mass models all assumed $f = 10^{-4}$
for BHL mass accretion, and because all started off with relatively
small masses, the amount of mass accreted during the $\sim 10^3$ yr
evolutions was negligible (Table 1); the largest increase was in model d, 
with initially 10 $M_\oplus$ protoplanets, where the amount of mass accreted
by one of the protoplanets was $\sim 0.002 M_\oplus$. 

\subsection{Giant-Planet Mass Protoplanets}

 We now turn to the models motivated by the Nice scenario for the
orbital evolution of the Solar System's giant planets (Tsiganis et al. 
2005; Gomes et al. 2005; Levison et al. 2008). We seek to learn what
might happen to massive gas and ice giant planets, formed by either
core accretion or disk instability, if they should encounter a
MGU disk phase, such as during an FU Orionis outburst.

 Table 2 summarizes the Nice scenario models with two different 
values of the BHL gas accretion factor $f$. In model M, with
$f = 10^{-4}$, an initially $1 M_{Jup}$ protoplanet gained
$0.6 M_{Jup}$ in mass during 3400 yr, yielding a mass accretion
rate of $\dot M \sim 2 \times 10^{-4} M_{Jup}$ yr$^{-1}$. In model 
Mh, with $f = 10^{-3}$, an initially $1 M_{Jup}$ protoplanet gained
$1.5 M_{Jup}$ in mass during 1300 yr, yielding a mass accretion
rate of $\sim 10^{-3} M_{Jup}$ yr$^{-1}$. The relatively high rate
of mass accretion in model Mh may not be physically reasonable.
Nelson \& Benz (2003) argued that such rates should be less than 
$\sim 10^{-4} M_{Jup}$ yr$^{-1}$. Kley (1999) found 
$\dot M = 4.35 \times 10^{-5} M_{Jup}$ yr$^{-1}$ in his standard model,
while Machida et al. (2010) found $\dot M \sim 10^{-5} M_{Jup}$ yr$^{-1}$ 
in their numerical simulations. However, in all of these 
other studies, the disks considered were not MGU disks, and
hence the protoplanetary mass accretion rates could be expected
to be significantly smaller. Nevertheless, the $\dot M$ estimates
for models M and Mh imply that the models with $f = 10^{-4}$
are probably more realistic than those with $f = 10^{-3}$.
The final masses listed in Table 2 show that the mass accretion
rates are systematically considerably higher in the $f = 10^{-3}$
models than in those with $f = 10^{-4}$, as is to be expected.
 
 Table 2 also shows that only 10
of the initial total of 32 protoplanets remained within the
disk during the evolutions: 22 hit either the inner or outer
boundary, 10 of the $f = 10^{-4}$ models, and 12 of the 
$f = 10^{-3}$ models. Considering the small number statistics,
there does not appear to be a strong dependence on the assumed
value of $f$. However, in the models with the most massive 
protoplanets (N, Nn), 7 out of 8 protoplanets were ejected, 
while 15 out of 24 were ejected in the lower mass models
(M, Mh, O, Oh, P, Ph). Apparently protoplanets with initial
masses of $\sim 3 M_{Jup}$ are more likely to undergo strong mutual
close encounters, which, coupled with the MGU disk perturbations,
eventually lead to their ejections. The higher overall ejection 
frequency for the Table 2 models compared to those in Table 1 is 
due in part to the longer time periods calculated for the 
Table 2 models, with the remaining difference being caused by
the stronger effects of close encounters between the much more massive
protoplanets in the Table 2 models.

 Figures 4 and 5 depict the semimajor axis evolutions of the
eight models listed in Table 2. As in the case of the $\sim$ Earth-mass
cores, the protoplanets are subjected to quasi-periodic perturbations
from the MGU disk's clumps and spiral arms, resulting in somewhat
chaotic orbital evolutions, but again without any clear evidence for
monotonically inward (or outward) migration. In general, the
semimajor axes of the surviving protoplanets remained in the
range of $\sim$ 5 AU to $\sim$ 15 AU, similar to the initial orbits,
in spite of mutual close encounters leading to frequent ejections
of the less fortunate, generally lower mass, protoplanets. 

 Table 3 summarizes the models that are the closest to the Nice model
for our Solar System, all calculated with the same value 
of $f = 10^{-4}$. The initial protoplanet masses in 
these models are closer to those of the current masses
of the giant planets in our Solar System than those of the previous
models. Model Q, in particular, has masses similar to those of
Jupiter, Saturn, Uranus, and Neptune, though the 0.1 $M_{Jup}$ 
protoplanets that represent the two ice giants start off with
masses about twice that of Uranus and Neptune. Model R explores
a situation with even more massive outer protoplanets, while
models S and T investigate a system where only Jupiter and Saturn
exist, starting at two different initial orbits for Saturn,
12 AU and 10 AU, respectively.

 Figures 6 and 7 display the evolutions of the semimajor axes
and eccentricities for the four models listed in Table 3. 
Model Q, the closest model to our outer Solar System, manages
to survive intact for a period of at least 1200 yr, though
only after undergoing a major orbital reshuffling: at the final
time shown in Figure 6, the initially innermost Jupiter-mass
protoplanet is now the outermost body, the initially Saturn-mass
protoplanet is the innermost body, and the two initially outer
ice giants are orbiting between the two gas giants, with all
four having semimajor axes between $\sim$ 9 AU and $\sim$ 13 AU.
Note the high eccentricities for prolonged periods for the two
ice giant-mass protoplanets in models Q (Figure 7),
implying that ejections are eventually likely for these bodies.
These systems would presumably become even more unstable once the 
disk gas is removed, with an uncertain final outcome. A similar
reordering of the orbital distribution occurs in model R, though 
in this case the initially 0.33 $M_{Jup}$ protoplanet, and one 
of the 0.5 $M_{Jup}$ protoplanets are ejected, leaving behind an
outer 1.4 $M_{Jup}$ gas giant and an inner 0.62 $M_{Jup}$ protoplanet.

 Model T shows that two gas giant planets can survive for $\sim$
4000 yr in a MGU disk, though they may well interchange their
orbital positions. On the other hand, with a different initial
orbital configuration, model S shows that the initially Saturn-mass
protoplanet might be ejected, leaving the initially Jupiter-mass
protoplanet as the sole survivor after 3800 yr. In both cases,
the eccentricities of the protoplanets tend to be modest (Figure 7c,d),
with $e \sim 0.05$ to 0.2. 

\section{Discussion}

 Tables 2 and 3 show that in a system with protoplanets of different
initial masses, in nearly every case the protoplanet that is
ejected is one of the lower mass protoplanets. The only exception
to this general rule was the initially 1 $M_{Jup}$ protoplanet in model 
Ph, which was ejected along with an initially 0.5 $M_{Jup}$ protoplanet.
Note, however, that even in this case, because of the high value
of $f = 10^{-3}$, the two other initially 0.5 $M_{Jup}$ protoplanets grew 
to masses of 2.6 $M_{Jup}$ and 1.6 $M_{Jup}$ by the end of the calculation,
and these were the bodies responsible for ejecting the initially 
1 $M_{Jup}$ protoplanet. Hence the general rule appears to be that the 
more massive protoplanets are left behind during close orbital encounters,
as is expected to be the case based on equipartition of energy:
the less massive body will receive a larger velocity perturbation
than the more massive body, and so is more likely to hit a disk boundary.

 We now briefly compare the results to those of the previous studies
of orbital migration in MGU disks. Boss (2005) considered the evolution
of fully three dimensional MGU disks with a mass of 0.091 $M_\odot$
extending from 4 AU to 20 AU around a 1 $M_\odot$ protostar, i.e., the
same situation as is considered here with multiple protoplanets
with varied masses. Jupiter-mass protoplanets inserted at 8 AU were 
found by Boss (2005) to orbit fairly stably, or to 
move out to $\sim$ 10 AU, over $\sim 10^3$ yr,
even while gaining mass by accretion. This implied that protoplanets 
in MGU disks do not immediately open disk gaps and undergo Type II 
migration. These results are quite consistent with the present models,
where the addition of other protoplanets does not prevent the
surviving protoplanets from orbiting relatively stably.

 Baruteau et al. (2011) considered the evolution
of two dimensional, thin MGU disks with a mass of 0.4 $M_\odot$
extending from 20 AU to 250 AU around a 1 $M_\odot$ protostar.
Saturn-mass and Jupiter-mass protoplanets inserted at 100 AU were found 
to migrate rapidly inward to $\sim$ 25 AU, on a time scale comparable 
to that expected for Type I migration, $\sim 10^4$ yr, while planets 
with $5 M_{Jup}$ migrated even faster, in $\sim 3 \times 10^3$ yr. 
Type II migration did not occur, as the planets were unable to
open disk gaps. The MGU nature of the evolving disk resulted in periodic
outward motions, rather than the monotonic inward motions of classic
Type I migration. These results are in basic agreement with the
present models. Baruteau et al. (2011) included the effects of the
planet's gravity on the disks, but did not include planet mass accretion (i.e., 
they fixed the planet masses), and their models were restricted to considering
the evolution of a single planet at a time, unlike the present models,
where planet-planet interactions are an important factor.
Most importantly, the Baruteau et al. (2011) models were limited to
orbits in the outer disk: their planets were forced to stop inward
migration at $\sim$ 25 AU, whereas the present models have protoplanets
that start inside 12 AU. The absence of inward migration
in the present models appears to be linked to the high inner disk 
temperatures ($\sim$ 600 K), leading to $Q >> 1$, and stifling to 
some extent the spiral arms just outside 4 AU (Figure 1), combined
with the chaotic outcomes of protoplanet interactions with a MGU disk.

 Michael et al. (2011) considered the evolution
of fully three dimensional MGU disks with a mass of 0.14 $M_\odot$
extending from 5 AU to 40 AU around a 1 $M_\odot$ protostar. Two
Jupiter-mass protoplanets inserted at 25 AU were found to migrate 
rapidly inward to $\sim$ 17 AU in $\sim 10^3$ yr, where both stalled.
The inward motion was again not monotonic, but rather jerky, due to
the MGU disk interactions. Similar to Baruteau et al. (2011), 
Michael et al. (2011) included the effects of the planet's gravity 
on the disk, did not include planetary mass accretion, and calculated
evolutions for only a single planet at a time. Michael et al. (2011)
also studied considerably larger disks than the present models,
but their inner disk boundary of 5 AU was quite similar to the 
4 AU value in the present models. The Michael et al. (2011) protoplanets
migrated inward but stopped at $\sim$ 17 AU, whereas in the present
models, the survivors clustered around distances of $\sim$ 8 AU
to $\sim$ 13 AU. This slight difference in outcomes can be 
attributed to the different underlying MGU disk structures used
in the two sets of models: Michael et al. (2011) started with a 
disk with a minimum $Q = 1.38$ at 26.7 AU, whereas in the present
models, the minimum $Q = 1.5$ occurred at the outer disk boundary
(20 AU), and rose to $Q > 10$ in the inner disk. Given these MGU disk
differences, the Michael et al. (2011) results seem to be compatible
with the present results.

 Finally, throughout the evolutions of all of these MGU disk models, 
disk mass flowed freely inward, past the orbiting protoplanets, 
and was accreted by the inner grid boundary at 4 AU. The total 
amount of mass accreted by the central region (i.e., the protostar) 
was typically $\sim 0.03 M_\odot$ (Tables 1, 2, 3) over a time 
period of $\sim$ 3000 yrs, leading to a protostellar mass 
accretion rate of $\sim 10^{-5} M_\odot$ yr$^{-1}$. This rate is comparable 
to the inferred mass accretion rates for T Tauri stars undergoing FU Orionis 
outbursts (e.g., Hartmann \& Kenyon 1996), confirming the applicability 
of these MGU disk models for protoplanetary systems undergoing
FU Orionis events. The Tables show that this central mass accretion
rate did not vary significantly across the models calculated,
showing that the varied protoplanet masses had little effect
on the overall evolution of the MGU disks.

\section{Conclusions}

 Given the limited number of models run, and the resulting highly 
incomplete examination of the initial conditions parameter space, 
one cannot draw too strong of a conclusion from these models,
but at the least, these models illustrate the range of outcomes
that could result from a MGU disk phase during planetary system
formation. Neverthless, it is clear that a MGU disk phase need
not be fatal to growing cores in the core accretion scenario, or
to giant planets formed by either core accretion or disk instability,
at least not for protoplanets with initial orbits in the range
of 6 AU to 12 AU from a solar-mass protostar. FU Orionis phases
thus need not be fatal to the giant planet formation process.
It is even conceivable that a Nice model-like scenario could be
constructed for protoplanets that survive a MGU disk phase,
though the most Nice-like model presented here (model Q) ended
up with Jupiter as the outermost body, rather than the innermost.
Other initial conditions might well lead to a more Nice-like outcome.

 The $\sim$ Earth-mass protoplanets are excited to relatively
high eccentricity orbits during the MGU disk phase, with 
$e \sim 0.2$ to 0.5. For the giant-planet-mass models, the final 
eccentricities are more modest, typically with $e \sim 0.05$ 
to 0.2, as is to be expected on the basis of equipartition
of energy. Hence, except for limited periods of time, the
orbital eccentricities for the giant planets are not as high as 
the highest values observed for Doppler-discovered extrasolar 
giant planets, where values as high as $e \sim 0.8$ have been
determined. However, most exoplanets have more modest eccentricities,
with over half having $e < 0.2$. The highest eccentricities found
for exoplanets presumably have their origins in planet-planet
scattering events, which can also result in significant orbital
inclinations, and even in retrograde orbital rotations, which seems
to be required in order to explain the orbital inclinations
deduced on the basis of the Rossiter-McLaughlin effect for
short-period exoplanets (e.g., Albrecht et al. 2012). 
The present models cannot address this possibility, as their
limitation to protoplanet orbits within the disk midplane
precludes interactions leading to inclined orbits.

 As Boss (2005) found, protoplanets located at $\sim$ 10 AU in a MGU 
disk can orbit relatively stably for significant periods of time 
($\sim 10^3$ yr or more), without undergoing monotonic inward 
Type-I-like migration, and without opening a disk gap, leading to
Type-II-like migration. Instead, the quasi-periodic gravitational
perturbations induced by the spiral arms and clumps of the MGU disk
result in eccentric orbits ($e \sim 0.2$), while close encounters
with the other protoplanets, combined with the MGU disk interactions,
can lead to a significant number of ``ejections'' of the less massive 
protoplanets through hitting the inner or outer disk boundaries, 
though these ``ejections'' might very well be ameliorated in models
that included a disk that extended from the true surface of the
protostar ($\sim$ 0.05 AU) out to much larger distances
($\sim$ 50 AU). Such improved models of protoplanet-MGU
disk interactions are needed to determine if observed exoplanets
on distant orbits (e.g., around HR 8799; Marois et al. 2008, 2010)
could have formed closer to their star and then been ejected outward,
or were formed more or less {\it in situ} (e.g., Boss 2011). 
Virtual protoplanet models based on the 60 AU-radius disk models of
Boss (2011) are in progress and will be the subject of a future
report.

 I thank Hal Levison for suggesting in 2007 the Nice model 
calculations at the Nobel Symposium \#135, John Chambers for
advice on orbital fitting procedures, Sandy Keiser for her 
invaluable assistance with the Carnegie Alpha Cluster, and the
referee for numerous suggestions for improving the paper. This research 
was supported in part by the NASA Origins of Solar Systems Program 
(NNX09AF62G) and contributed in part to the NASA Astrobiology Institute 
(NNA09DA81A). The calculations were performed on the Carnegie Alpha Cluster, 
the purchase of which was partially supported by NSF Major Research
Instrumentation grant MRI-9976645.

\clearpage
\begin{deluxetable}{ccccccc}
\tablecaption{Initial conditions and final outcomes for the $\sim$ Earth-mass
protoplanets embedded in MGU disks with identical $f$ BHL factors. $M_i$ and 
$a_i$ are the initial planet mass and orbital radius, $M_f$ is the planet mass 
at the final time $t_f$, and $\Delta M_s/M_\odot$ is the amount of disk mass 
accreted by the central protostar by the end of the evolution. \label{tbl-1}}
\tablehead{\colhead{model} & 
\colhead{\quad $f$ \quad } & 
\colhead{\quad $M_i/M_\oplus$ \quad } & 
\colhead{\quad $a_i$ (AU) \quad } & 
\colhead{\quad $t_f$ (yr) \quad } & 
\colhead{\quad $M_f/M_\oplus$ \quad } & 
\colhead{\quad $\Delta M_s/M_\odot$ \quad } }
\startdata

a  & $10^{-4}$  &  0.01  &  6  & 730  &  0.01      & 0.023 \\
   &   "        &  0.01  &  8  & "    &  0.01      & " \\
   &   "        &  0.01  & 10  & "    &  eject out & " \\
   &   "        &  0.01  & 12  & "    &  0.01      & " \\

b  & $10^{-4}$  &  0.1   &  6  & 900  &  eject in  & 0.029 \\
   &   "        &  0.1   &  8  & "    &  eject in  & " \\   
   &   "        &  0.1   & 10  & "    &  0.1       & " \\
   &   "        &  0.1   & 12  & "    &  0.1       & " \\

c  & $10^{-4}$  &  1.0   &  6  & 730  &  1.0       & 0.025 \\
   &   "        &  1.0   &  8  & "    &  eject in  & " \\
   &   "        &  1.0   & 10  & "    &  eject out & " \\
   &   "        &  1.0   & 12  & "    &  1.0       & " \\

d  & $10^{-4}$  &  10.0  &  6  & 760  &  10.0      & 0.025 \\
   &   "        &  10.0  &  8  & "    &  eject out & " \\
   &   "        &  10.0  & 10  & "    &  eject out & " \\
   &   "        &  10.0  & 12  & "    &  10.0      & " \\

\enddata
\end{deluxetable}
\clearpage

\suppressfloats

\suppressfloats

\begin{deluxetable}{ccccccc}
\tablecaption{Initial conditions and final outcomes for the giant 
protoplanets embedded in MGU disks with varied $f$ BHL factors. \label{tbl-2}}
\tablehead{\colhead{model} & 
\colhead{\quad $f$ \quad } & 
\colhead{\quad $M_i/M_{Jup}$ \quad } & 
\colhead{\quad $a_i$ (AU) \quad } & 
\colhead{\quad $t_f$ (yr) \quad } & 
\colhead{\quad $M_f/M_{Jup}$ \quad } &
\colhead{\quad $\Delta M_s/M_\odot$ \quad } }

\startdata
M  & $10^{-4}$  &  1.0  &  6  & 3400 & 1.6       & 0.028 \\
   &   "        &  0.33 &  8  & "    & eject out & " \\
   &   "        &  0.33 & 10  & "    & eject out & " \\
   &   "        &  0.33 & 12  & "    & eject in  & " \\
Mh & $10^{-3}$  &  1.0  &  6  & 1300 & 2.5       & 0.023 \\
   &   "        &  0.33 &  8  & "    & eject out & " \\
   &   "        &  0.33 & 10  & "    & eject out & " \\
   &   "        &  0.33 & 12  & "    & eject out & " \\
N  & $10^{-4}$  &  1.0  &  6  & 3400 & eject out & 0.042 \\
   &   "        &  3.0  &  8  & "    & 4.0       & " \\
   &   "        &  3.0  & 10  & "    & eject out & " \\
   &   "        &  3.0  & 12  & "    & eject out & " \\
Nh & $10^{-3}$  &  1.0  &  6  & 410  & eject out & 0.019 \\
   &   "        &  3.0  &  8  & "    & eject out & " \\
   &   "        &  3.0  & 10  & "    & eject in  & " \\
   &   "        &  3.0  & 12  & "    & eject in  & " \\
O  & $10^{-4}$  &  1.0  &  6  & 3050 & eject out & 0.035 \\
   &   "        &  1.0  &  8  & "    & eject in  & " \\
   &   "        &  1.0  & 10  & "    & 1.5       & " \\
   &   "        &  1.0  & 12  & "    & 1.1       & " \\
Oh & $10^{-3}$  &  1.0  &  6  & 1080 & eject out & 0.023 \\
   &   "        &  1.0  &  8  & "    & eject in  & " \\
   &   "        &  1.0  & 10  & "    & 3.1       & " \\
   &   "        &  1.0  & 12  & "    & eject out & " \\
P  & $10^{-4}$  &  1.0  &  6  & 3050 & 1.4       & 0.029 \\
   &   "        &  0.5  &  8  & "    & eject out & " \\
   &   "        &  0.5  & 10  & "    & eject out & " \\
   &   "        &  0.5  & 12  & "    & 0.91      & " \\
Ph & $10^{-3}$  &  1.0  &  6  & 1500 & eject out & 0.024 \\
   &   "        &  0.50 &  8  & "    & eject out & " \\
   &   "        &  0.50 & 10  & "    & 2.6       & " \\
   &   "        &  0.50 & 12  & "    & 1.6       & " \\
\enddata
\end{deluxetable}

\suppressfloats

\clearpage
\begin{deluxetable}{ccccccc}
\tablecaption{Initial conditions and final outcomes for the giant 
protoplanets embedded in MGU disks with identical $f$ BHL factors. \label{tbl-3}}
\tablehead{\colhead{model} & 
\colhead{\quad $f$ \quad } & 
\colhead{\quad $M_i/M_{Jup}$ \quad } & 
\colhead{\quad $a_i$ (AU) \quad } & 
\colhead{\quad $t_f$ (yr) \quad } & 
\colhead{\quad $M_f/M_{Jup}$ \quad } &
\colhead{\quad $\Delta M_s/M_\odot$ \quad } }

\startdata
Q  & $10^{-4}$  &  1.0  &  6  & 1200 & 1.1       & 0.024 \\
   &   "        &  0.33 &  8  & "    & 0.37      & " \\
   &   "        &  0.10 & 10  & "    & 0.10      & " \\
   &   "        &  0.10 & 12  & "    & 0.10      & " \\
R  & $10^{-4}$  &  1.0  &  6  & 1200 & 1.4       & 0.022 \\
   &   "        &  0.33 &  8  & "    & 0.62      & " \\
   &   "        &  0.50 & 10  & "    & eject in  & " \\
   &   "        &  0.50 & 12  & "    & eject in  & " \\
S  & $10^{-4}$  &  1.0  &  6  & 3800 & 1.3       & 0.031 \\
   &   "        &  0.33 & 12  & "    & eject in  & " \\
T  & $10^{-4}$  &  1.0  &  6  & 3800 & 1.3       & 0.029 \\
   &   "        &  0.33 & 10  & "    & 0.40      & " \\
\enddata
\end{deluxetable}
\clearpage

\suppressfloats

\begin{figure}
\vspace{-1.0in}
\caption{[Removed to fit within the limits for submission.] Midplane density contours for the MGU disk at the phase
when the protoplanets (red circles) are first inserted into the models.
The disk starts with a mass of $0.091 M_\odot$, with an outer radius of 
20 AU and an inner radius of 4 AU, through which mass accretes onto the 
initially 1 $M_\odot$ central protostar. Density contours are shown
in g cm$^{-3}$. Red circles denote the locations where protoplanets
are inserted, at distances of 6, 8, 10, or 12 AU from the protostar,
starting at 9 o'clock and rotating counterclockwise, respectively.}
\end{figure}

\clearpage

\begin{figure}
\vspace{0.0in}
\plotone{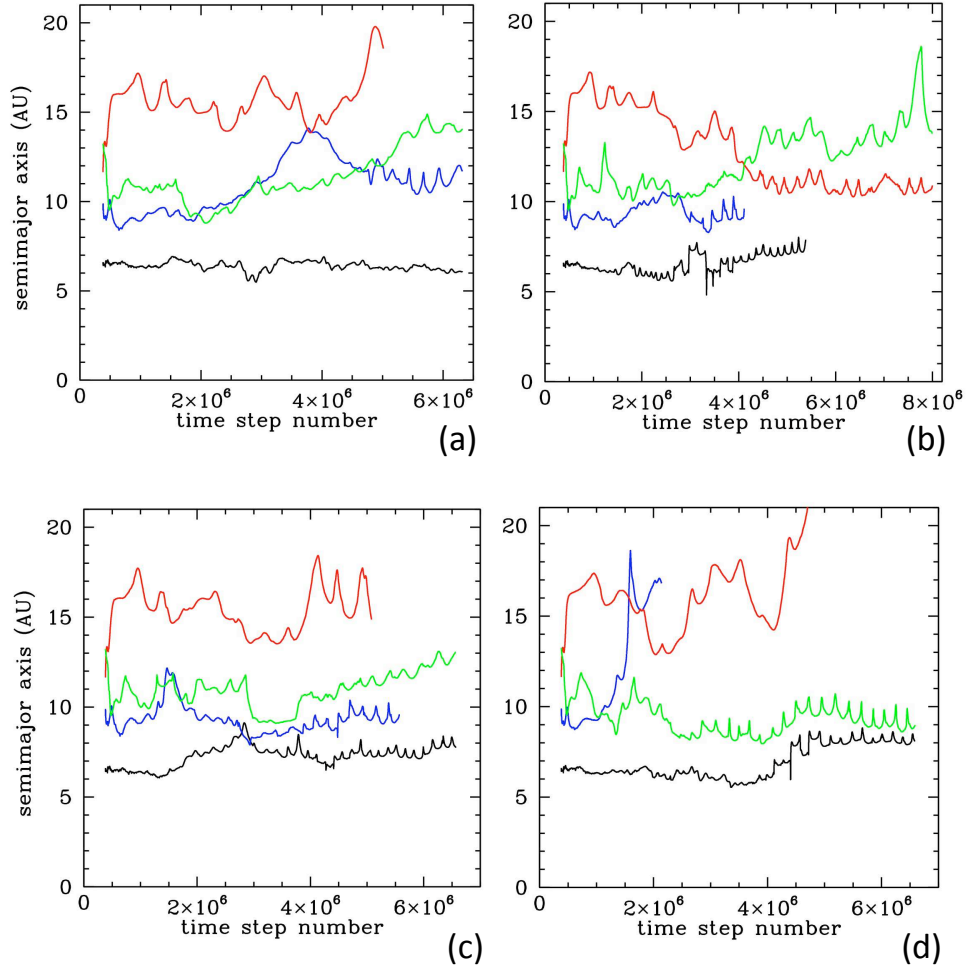}
\vspace{-2.5in}
\caption{Time evolution of the semimajor axes of $\sim$ Earth-mass
embedded protoplanets with initial masses of (a) 0.01 $M_\oplus$ (model 
a), (b) 0.10 $M_\oplus$ (model b), (c) 1.00 $M_\oplus$ (model c), and
(d) 10.0 $M_\oplus$ (model d). Elapsed times since protoplanet insertion 
for each model are: (a) 730 yr, (b) 930 yr, (c) 730 yr, and (d) 760 yr.
Data are plotted every 100 time steps.
Protoplanets were inserted at radii of 6 AU (black), 8 AU (blue),
10 AU (red), or 12 AU (green), as shown in Figure 1. Protoplanets that
strike the inner (4 AU) or outer (20 AU) disk boundaries are considered
to be ejected and are dropped from the calculations.}
\end{figure}

\clearpage

\begin{figure}
\vspace{0.0in}
\plotone{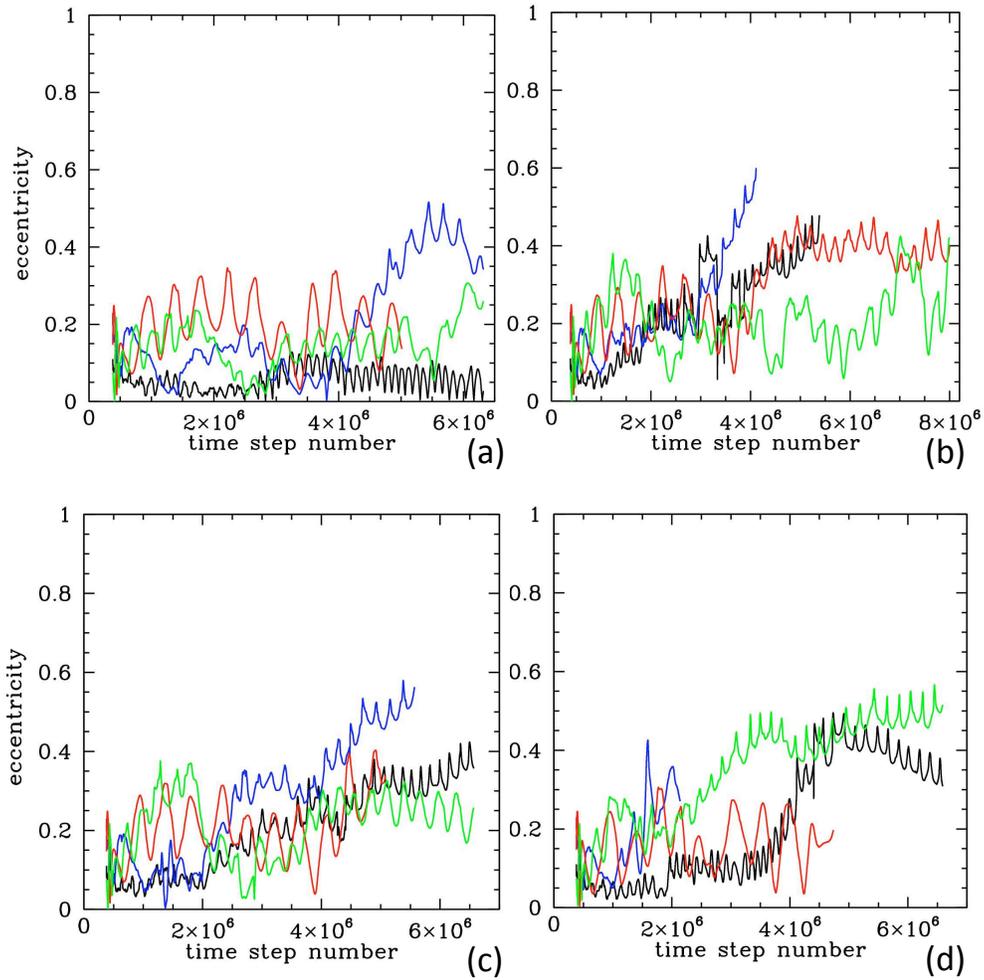}
\vspace{-2.5in}
\caption{Time evolution of the eccentricities of $\sim$ Earth-mass
embedded protoplanets, plotted in the same manner as in Figure 2, 
for models a, b, c, and d.}
\end{figure}

\clearpage

\begin{figure}
\vspace{0.0in}
\plotone{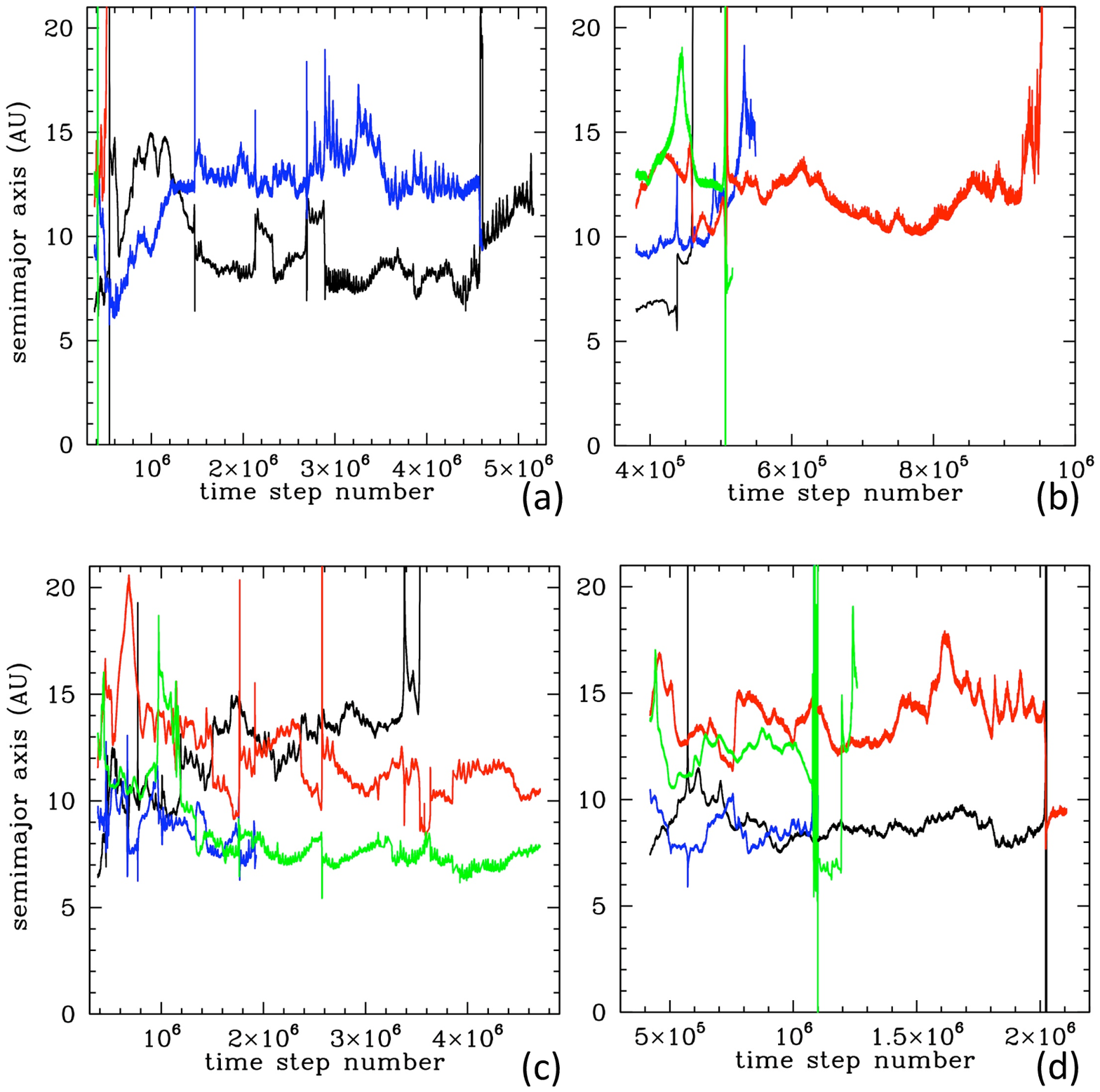}
\vspace{-2.5in}
\caption{Time evolution of the semimajor axes of giant planet-mass embedded 
protoplanets, plotted as in Figure 2. 
Elapsed times: (a) model N: 3400 yr, (b) model Nh: 410 yr, 
(c) model O: 3050 yr, and (d) model Oh: 1080 yr.}
\end{figure}

\clearpage

\begin{figure}
\vspace{0.0in}
\plotone{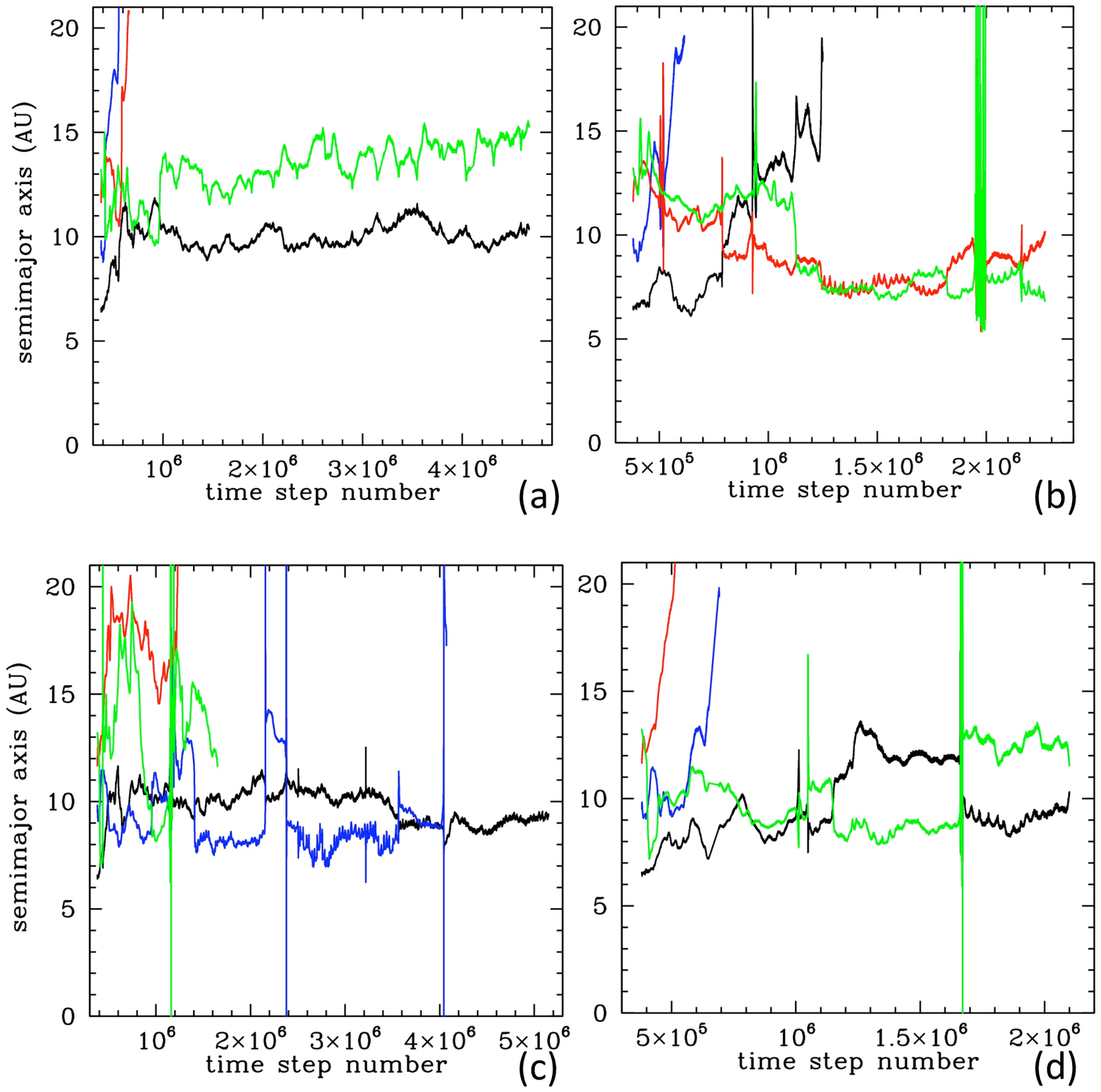}
\vspace{-2.5in}
\caption{Time evolution of the semimajor axes of giant planet-mass embedded 
protoplanets, plotted as in Figure 2.
Elapsed times: (a) model P: 3050 yr, (b) model Ph: 1500 yr, 
(c) model M: 3400 yr, and (d) model Mh: 1300 yr.}
\end{figure}

\clearpage

\begin{figure}
\vspace{0.0in}
\plotone{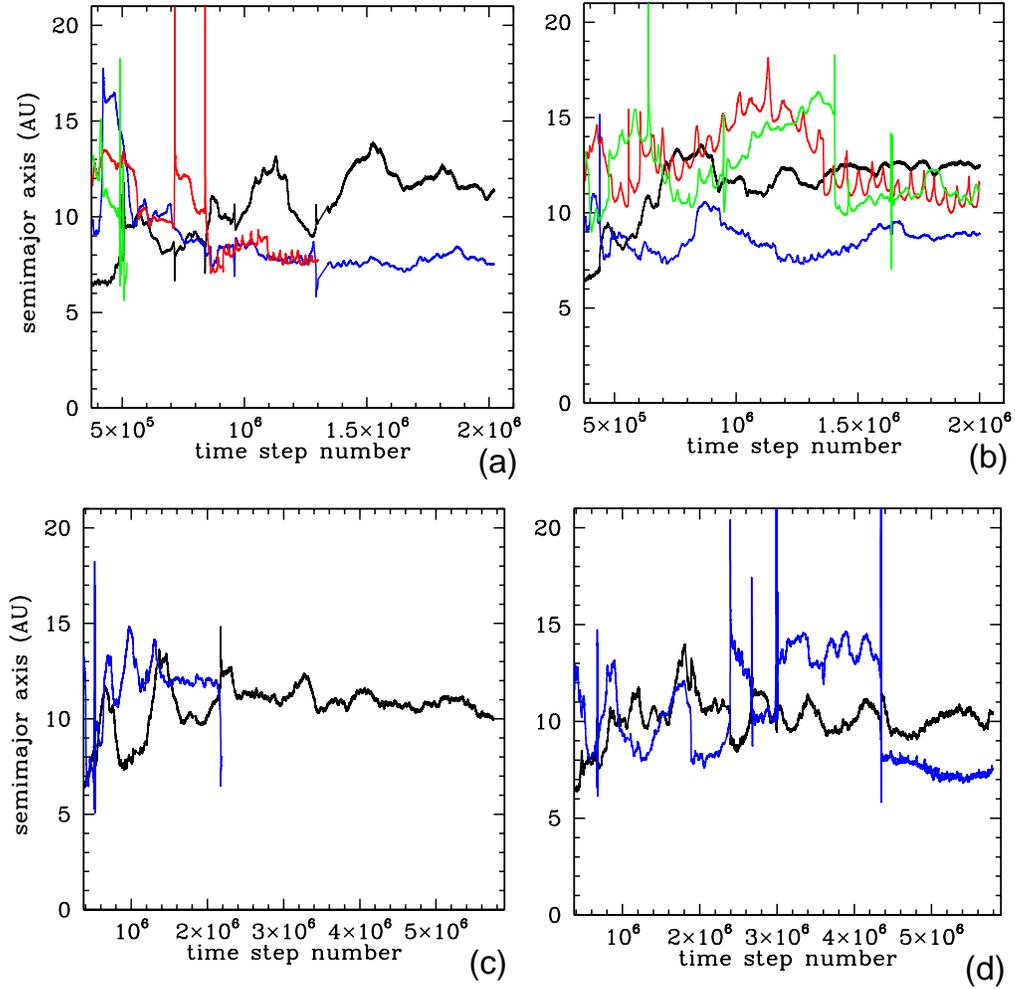}
\vspace{-2.5in}
\caption{Time evolution of the semimajor axes of giant planet-mass
embedded protoplanets, plotted as in Figure 2. Elapsed times:
(a) model R: 1200 yr, (b) model Q: 1200 yr, (c) model S: 3800 yr, and 
(d) model T: 3800 yr.}
\end{figure}

\clearpage

\begin{figure}
\vspace{0.0in}
\plotone{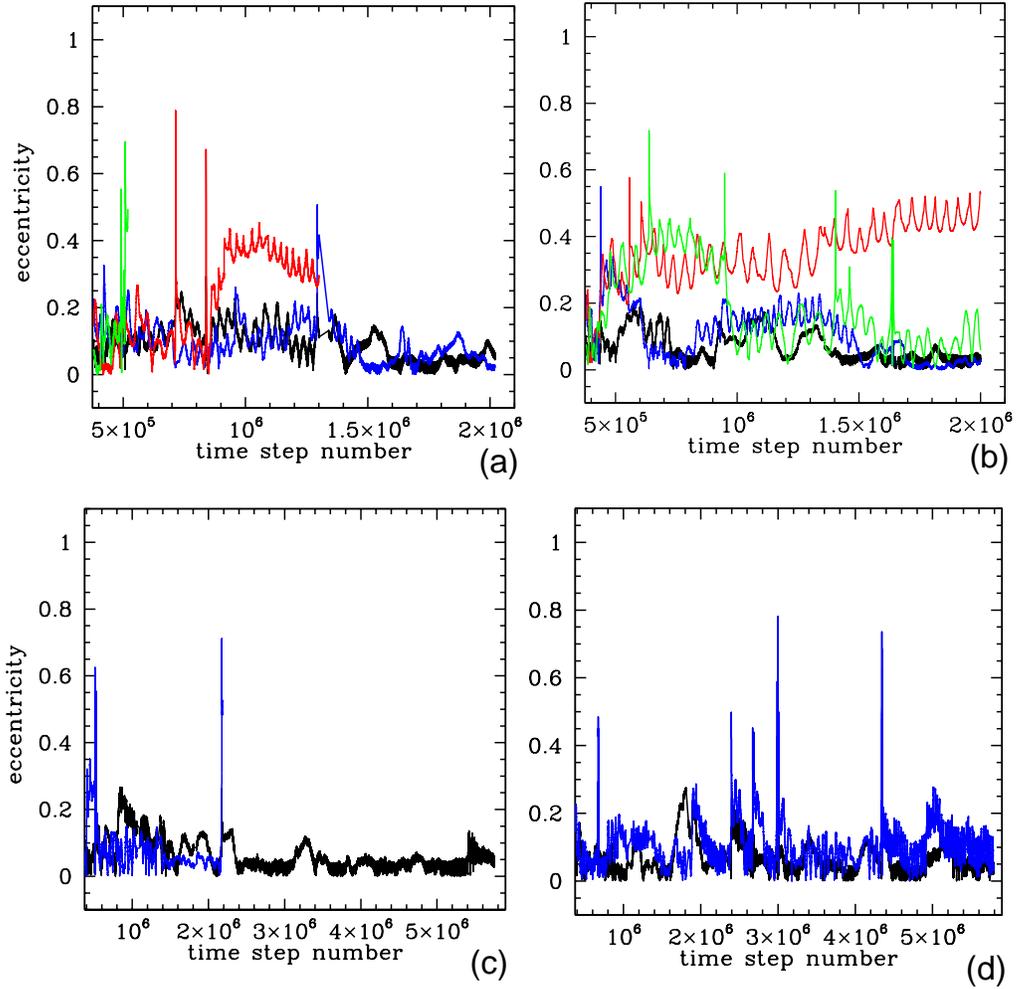}
\vspace{-2.5in}
\caption{Time evolution of the eccentricities of giant planet-mass
embedded protoplanets, plotted as in Figure 6. Elapsed times:
(a) model R: 1200 yr, (b) model Q: 1200 yr, (c) model S: 3800 yr, and 
(d) model T: 3800 yr.}
\end{figure}

\enddocument